\documentclass[twocolumn,showpacs,preprintnumbers,amsmath,amssymb]{revtex4}

\usepackage{graphicx}
\usepackage{dcolumn}
\usepackage{bm}

\newcommand{\bq}{\begin{equation}}
\newcommand{\eq}{\end{equation}}
\newcommand{\bqa}{\begin{eqnarray}}
\newcommand{\eqa}{\end{eqnarray}}
\newcommand{\nn}{\nonumber \\}

\def\be     {\begin{equation}}
\def\ee     {\end{equation}}
\def\bea        {\begin{eqnarray}}
\def\eea        {\end{eqnarray}}
\def\bnn    {\begin{eqnarray*}}
\def\enn    {\end{eqnarray*}}

\begin{document}

\title{Role of doped holes in a U(1) spin liquid}
\author{Ki-Seok Kim }
\affiliation{Korea Institute for Advanced Study, Seoul 130-012,
Korea}
\date{\today}

\begin{abstract}
In the context of the SU(2) slave boson theory we show that
condensation of holons can result in the zero mode of a nodal
spinon in a single instanton potential. Instanton contribution in
the presence of the zero mode induces the 't Hooft effective
interaction, here mass to the spinon. We find that the spinon mass
is determined by the state of instantons in the presence of the
zero mode. The mass corresponds to antiferromagnetic moment of the
nodal spinon. Considering the state of instantons, we discuss the
possibility of coexistence between antiferromagnetism and $d-wave$
superconductivity in underdoped cuprates.
\end{abstract}


\maketitle

High $T_c$ superconductivity ($SC$) is believed to result from
hole doping to an antiferromagnetic Mott insulator ($AFMI$). Hole
doping to the $AFMI$ destroys the antiferromagnetic long range
order and causes a paramagnetic Mott insulator ($PMMI$) usually
dubbed the pseudogap phase. High $T_c$ $SC$ is expected to occur
by further hole doping to the $PMMI$\cite{Senthil_Lee}. Recently,
the $PMMI$ is proposed to be the U(1) spin liquid ($U1SL$)
described by $QED_3$ in terms of massless Dirac spinons
interacting via non-compact U(1) gauge fields, which originates
from irrelevance of instanton excitations of compact U(1) gauge
fields in the large flavor limit\cite{Senthil_Lee,U1SL}. According
to this scenario, high $T_c$ $SC$ arises from hole doping to the
$U1SL$.

In the present study we investigate the role of hole doping in the
$U1SL$. In the context of the SU(2) slave boson
theory\cite{SU2_gauge} doped holes are represented by SU(2) holon
doublets. It is the key observation that isospin interactions
between spinons and holons can appear in the SU(2) slave boson
theory. This new interaction is shown to result in the zero mode
of a nodal spinon in a single instanton potential\cite{Jackiw},
which appears in the $SC$ state resulting from the condensation of
holon doublets. In high energy physics the instanton contribution
in the presence of the fermion zero mode is well known to induce
the 't Hooft effective interaction, here mass to the
spinon\cite{tHooft,Dmitri}. We find that the spinon mass is
determined by the state of instantons in the presence of the zero
mode. The mass corresponds to antiferromagnetic moment of the
nodal spinon\cite{Marston,Don_Kim}. Considering the state of
instantons, we discuss the possibility of coexistence between
antiferromagnetism ($AF$) and $d-wave$ $SC$ in underdoped
cuprates.

We consider an effective Lagrangian describing hole doped $U1SL$
in the context of the SU(2) slave boson
theory\cite{Senthil_Lee,SU2_gauge} \bqa && Z =
\int{D\psi_n}{Dz_n}{Da_{\mu}}e^{-\int{d^3x} {\cal L}} , \nn &&
{\cal L} = \sum_{n,m = 1}^{2} \Bigl[
\bar{\psi}_{n}\gamma_{\mu}(\partial_{\mu}\delta_{nm} +
ia_{\mu}\tau^{3}_{nm})\psi_{m} \nn && +
|(\partial_{\mu}\delta_{nm} + ia_{\mu}\tau^{3}_{nm} )z_{m}|^2 +
m^{2}|z_{n}|^{2} + \frac{u}{2}|z_{n}|^{4}  \nn && +
\frac{1}{2}\sum_{n', m' = 1}^{2}
G\bar{\psi}_{n}\tau^{k}_{nm}\psi_{m}
z^{\dagger}_{n'}\tau^{k}_{n'm'}z_{m'} \Bigr] +
\frac{1}{2e^2}|\partial\times{a}|^2 . \eqa Here $\psi_{n} = \left(
\begin{array}{c} \chi^{+}_{n} \\  \chi^{-}_{n} \end{array} \right)$
is the four component massless Dirac fermion where $n = 1, 2$
represent SU(2) isospin indices. The two component spinors
$\chi^{\pm}_{n}$ are given by $\chi^{+}_{1} = \left(
\begin{array}{c} f_{1e\uparrow} \\ f_{1o\uparrow} \end{array}
\right)$, $\chi^{-}_{1} = \left( \begin{array}{c} f_{2o\uparrow}
\\ f_{2e\uparrow} \end{array} \right)$,
$\chi^{+}_{2} = \left( \begin{array}{c} f_{1e\downarrow}^{\dagger}
\\ f_{1o\downarrow}^{\dagger} \end{array} \right)$, and
$\chi^{-}_{2} = \left( \begin{array}{c} f_{2o\downarrow}^{\dagger} \\
f_{2e\downarrow}^{\dagger} \end{array} \right)$, respectively. In
the spinon field $f_{abc}$ $a = 1, 2$ represent the nodal points
of $(\pi/2,\pi/2)$ and $(-\pi/2,\pi/2)$, $b = e, o$, even and odd
sites, and $c = \uparrow, \downarrow$, its spin,
respectively\cite{Don_Kim}. The Dirac matrices $\gamma_{\mu}$ are
given by $\gamma_{0} = \left( \begin{array}{cc} \sigma_{3} & 0 \\
0 & -\sigma_{3} \end{array} \right)$, $\gamma_{1} = \left(
\begin{array}{cc} \sigma_{2} & 0 \\ 0 & -\sigma_{2} \end{array} \right)$,
and $\gamma_{2} = \left( \begin{array}{cc} \sigma_{1} & 0 \\ 0 &
-\sigma_{1} \end{array} \right)$, respectively, where they satisfy
the Clifford algebra $[\gamma_{\mu},\gamma_{\nu}]_{+} =
2\delta_{\mu\nu}$\cite{Don_Kim}. $z_{n}$ represents the SU(2)
holon doublet with the isospin indices $n = 1, 2$\cite{SU2_gauge}.
$m$ and $u$ denote the mass and self-interaction of the holon,
respectively. We model an effective holon potential with easy
plane anisotropy resulting from the contribution of high energy
fermions\cite{SU2_gauge}. A coupling between the spinon and holon
isospins originates from gauge interactions mediated by the time
component of SU(2) gauge fields\cite{SU2_gauge}. Similar
consideration can be found in Ref. \cite{SU2_gauge}. $G$ denotes
the coupling strength between the isospins and $\tau^k$ acts on
the SU(2) isospin space. As will be seen below, this isospin
interaction plays an important role on instanton excitations in
the $SC$ state. $a_{\mu}$ is a compact U(1) gauge field in itself.
The kinetic energy of the gauge field arises from particle-hole
excitations of high energy quasiparticles\cite{Compactness}. $e$
is an effective internal charge, not a real electric charge.

In passing, we discuss an effective field theory in the $SC$ state
of Eq. (1) without the isospin interaction. Holon condensation
$<z_{1(2)}> \not= 0$ results in the $SC$ state, causing the U(1)
gauge field $a_{\mu}$ to be massive via the Anderson-Higgs
mechanism. Integration over the massive gauge field gives an
effective field theory in terms of electrons $c_{n} =
z^{\dagger}_{n}\psi_{n}$ and holon pairs $z_{1}z_{2}$. The spinons
and holons are confined to form internal charge neutral objects.
This phase can be considered to be the Higgs-confinement phase in
the context of the gauge theory\cite{Fradkin}. In the easy plane
limit $z_{n} = e^{i\phi_{n}}$, the low energy effective Lagrangian
is given by \bqa && {\cal L} = \frac{\rho}{2}|\partial_{\mu}\phi -
2A_{\mu}|^{2} + \frac{i}{2}(\partial_{\mu}\phi -
2A_{\mu})\bar{c}_{n}\gamma_{\mu}c_{n} \nn && +
\bar{c}_{n}\gamma_{\mu}(\partial_{\mu} + iA_{\mu})c_{n} +
\frac{1}{2g}|\bar{c}_{1}\gamma_{\mu}c_{1} -
\bar{c}_{2}\gamma_{\mu}c_{2}|^{2}  .  \eqa Here $\phi$ is the
phase field of the holon pair, $e^{i\phi} = z_{1}z_{2} =
e^{i(\phi_{1} + \phi_{2})}$. An external electromagnetic field
$A_{\mu}$ is introduced. $\rho$ is the stiffness parameter
proportional to hole concentration $\delta$ and $1/g \sim 1/\rho$,
the strength of four fermion interaction. Surprisingly, this
effective Lagrangian is nothing but that of the $d-wave$ $BCS$
superconductor\cite{DHLee}. A detailed discussion of this theory
can be found in Refs. \cite{DHLee,Kim_KT}. In this letter we show
that the presence of the isospin interaction can alter this
effective field theory completely.

Separating the compact U(1) gauge field $a_{\mu}$ into $a_{\mu} =
a_{\mu}^{cl} + a_{\mu}^{qu}$ where $a_{\mu}^{cl}$ represents an
instanton configuration and $a_{\mu}^{qu}$, gaussian fluctuations,
and integrating over the massless Dirac spinon field in Eq. (1),
we obtain a fermion determinant including the isospin interaction.
In order to calculate the determinant we solve an equation of
motion in the presence of a single monopole potential
$a_{\mu}^{cl} =
a(r)\epsilon_{3\alpha\mu}x_{\alpha}$\cite{Instanton_monopole} and
its corresponding hedgehog configuration $I_{\mu}^{int} =
\frac{1}{2}z^{\dagger}_{n}\tau^{\mu}_{nm}z_{m} = \Phi(r)x_{\mu}$
where $a(r)$ and $\Phi(r)$ are proportional to $r^{-2}$ in $r
\rightarrow \infty$ with $r = \sqrt{\tau^{2} + x^2 +
y^2}$\cite{Jackiw,Potential} \bqa &&
(\gamma_{\mu}\partial_{\mu}\delta_{nm} -
ia(r)(\gamma\times{x})_{3}\tau^{3}_{nm} +
G\Phi(r){x}_{\mu}\tau^{\mu}_{nm})\psi_{m} \nn && = E\psi_{n} .
\eqa In the absence of the isospin interaction it is well known
that there are no fermion zero modes\cite{Marston}. On the other
hand, the presence of the isospin interaction results in a fermion
zero mode. In the SU(2) gauge theory of massless Dirac fermions
and adjoint Higgs fields interacting via SU(2) gauge fields,
Jackiw and Rebbi showed that a Dirac equation coupled to the
isospin of the Higgs field has a fermion zero mode in a single
magnetic monopole potential\cite{Jackiw}. Following Jackiw and
Rebbi, we show that Eq. (3) also has a zero mode. We rewrite Eq.
(3) in terms of the two component spinors $\chi^{\pm}_{n}$ with $E
= 0$ \bqa && (\sigma_{3}\partial_{\tau})_{ij}\chi^{\pm}_{jn} +
(\sigma_{2}\partial_{x})_{ij}\chi^{\pm}_{jn} +
(\sigma_{1}\partial_{y})_{ij}\chi^{\pm}_{jn} \nn && -
iay(\sigma_{2})_{ij}\chi^{\pm}_{jm}(\tau^{3T})_{mn} +
iax(\sigma_{1})_{ij}\chi^{\pm}_{jm}(\tau^{3T})_{mn} \nn &&  \pm
G\Phi{x}_{\mu}\chi^{\pm}_{jm}(\tau^{\mu{T}})_{mn}  = 0 . \eqa
Inserting $\chi^{\pm}_{in} = \mathcal{M}^{\pm}_{im}\tau^{2}_{mn}$
with a two-by-two matrix $\mathcal{M}^{\pm}$ into the above, we
obtain \bqa && \sigma_{3}\partial_{\tau}\mathcal{M}^{\pm} +
\sigma_{2}\partial_{x}\mathcal{M}^{\pm} +
\sigma_{1}\partial_{y}\mathcal{M}^{\pm} \nn && +
iay\sigma_{2}\mathcal{M}^{\pm}\sigma^{3} -
iax\sigma_{1}\mathcal{M}^{\pm}\sigma^{3} \mp
G\Phi\mathcal{M}^{\pm}{x}_{\mu}\sigma^{\mu} = 0 . \eqa Now the
isospin matrices and the Dirac matrices are
indistinguishable\cite{Jackiw}. Finally, representing the matrix
$\mathcal{M}^{\pm}$ in $\mathcal{M}^{\pm}_{im} =
g^{\pm}\delta_{im} + g_{\mu}^{\pm}\sigma^{\mu}_{im}$, we obtain
coupled equations of motion for the numbers $g^{\pm}$ and
$g_{\mu}^{\pm}$ \bqa && (\partial_{\tau} \mp G\Phi\tau){g}^{\pm} -
i(\partial_{x} + ax \pm G\Phi{y})g^{\pm}_{1} \nn && +
i(\partial_{y} + ay \pm G\Phi{x} )g^{\pm}_{2} = 0 , \nn &&
(\partial_{x} - ax \mp G\Phi{y})g^{\pm} + i(\partial_{\tau} \pm
G\Phi\tau)g^{\pm}_{1} \nn && - i(\partial_{y} - ay \pm G\Phi{x}
)g^{\pm}_{3} = 0 , \nn && (\partial_{y} - ay \mp G\Phi{x})g^{\pm}
- i(\partial_{\tau} \pm G\Phi\tau)g^{\pm}_{2} \nn && +
i(\partial_{x} - ax \pm G\Phi{y})g^{\pm}_{3} = 0 , \nn &&
(\partial_{\tau} \mp G\Phi\tau)g^{\pm}_{3} + (\partial_{x} + ax
\mp G\Phi{y} )g^{\pm}_{2} \nn && + (\partial_{y} + ay \mp G\Phi{x}
)g^{\pm}_{1} = 0 . \eqa These equations yield the following zero
mode equations \bqa && (\partial_{\tau} + G\Phi\tau){g}^{-} = 0 ,
\nn && (\partial_{x} - ax + G\Phi{y})g^{-} = 0 , \nn &&
(\partial_{y} - ay + G\Phi{x})g^{-} = 0 . \eqa The zero mode
solution $g^{-}$ is given by $g^{-} \sim
exp\Bigl[-\int{d\tau}G\Phi(r)\tau + \int{dx}(a(r)x - G\Phi(r)y) +
\int{dy}(a(r)y - G\Phi(r)x)\Bigr]$. Without the isospin
interaction it can be easily seen that there exist no normalizable
fermion zero modes\cite{Marston}. The existence of the zero mode
makes the fermion determinant zero in the single monopole
excitation. As a result the condensation of magnetic monopoles is
forbidden. It is well known that the monopole condensation causes
confinement of charged particles\cite{U1SL,Polyakov}. The
suppression of monopole condensation results in
deconfinement\cite{Log_confinement} of internally charged
particles, here the spinons and holons. This deconfined $SC$ state
completely differs from the usual one corresponding to the
Higgs-confinement phase described by Eq. (2). Below we discuss an
effective field theory to describe this unusual $SC$ state.

In high energy physics it is well known that the instanton
contribution in the presence of the fermion zero mode gives rise
to an effective interaction to the fermions\cite{tHooft,Dmitri}.
This interaction is usually called the 't Hooft effective
interaction. In order to obtain the effective fermion interaction
it is necessary to average the partition function in Eq. (1) over
various instanton and anti-instanton configurations. Following
Ref. \cite{Dmitri}, first we consider a partition function in a
single instanton potential \bqa && Z_{\psi} =
\int{D\psi_{n}}e^{-\int{d^3x}
\bar{\psi}_{n}\gamma_{\mu}\partial_{\mu}\psi_{n}}\Bigl(m -
V^{I}[\psi_{n}]\Bigr) , \nn && V^{I}[\psi_{n}] = \int{d^3x}\Bigl(
\bar{\psi}_{n}(x)\gamma_{\mu}\partial_{\mu}\Phi^{I}_{n}(x)\Bigr)
\nn && \times
\int{d^3y}\Bigl(\bar{\Phi}_{n}^{I}(y)\gamma_{\mu}\partial_{\mu}\psi_{n}(y)\Bigr)
. \eqa Here $\Phi_{n}^{I}$ is the zero mode obtained from Eq. (7).
A fermion mass $m$ is introduced. Later the chiral limit $m
\rightarrow 0$ will be chosen. The effective action including the
effective potential $V^{I}[\psi_{n}]$ in Eq. (8) gives a correct
green function in a single instanton potential\cite{Dmitri},
$S^{I}(x,y) = <\psi_{n}(x)\bar{\psi}_{n}(y)> = -
\frac{\Phi_{n}^{I}(x)\bar\Phi_{n}^{I}(y)}{m} + S_{0}(x,y)$ with
the bare propagator $S_{0}(x,y) =
(\gamma_{\mu}\partial_{\mu})^{-1}\delta(x-y)$. Thus the partition
function Eq. (8) can be used for instanton average\cite{Dmitri}.
The partition function in the presence of $N_{+}$ instantons and
$N_{-}$ anti-instantons can be easily built up\cite{Dmitri} \bqa
&& Z_{\psi} =
\int{D\psi_{n}}{Da^{qu}_{\mu}}e^{-\int{d^3x}\bar{\psi}_{n}\gamma_{\mu}(\partial_{\mu}
- ia^{qu}_{\mu})\psi_{n}} \nn && \times   \Bigl(m -
<V^{I}[\psi_{n}]>\Bigr)^{N_+}\Bigl(m - <V^{I}[\psi_{n}]>
\Bigr)^{N_-} .\eqa Here we admit a non-compact U(1) gauge field
$a^{qu}_{\mu}$ representing gaussian fluctuations. Below the index
$qu$ is omitted. $<...>$ means averaging over the individual
instantons. Introducing instanton averaged non-local fermion
vertices $Y_{\pm} = - V<V^{I}[\psi_{n}]>
=-\int{d^3z_{I(\bar{I})}}V^{I(\bar{I})}[\psi_{n}]$ with volume
$V$, where $z_{I(\bar{I})}$ represent instanton center
positions\cite{Dmitri}, we obtain a partition function in the
chiral limit $m \rightarrow 0$ \bqa && Z_{\psi} =
\int{D\psi_{n}}{Da_{\mu}}
e^{-\int{d^3x}\bar{\psi}_{n}\gamma_{\mu}(\partial_{\mu} -
ia_{\mu})\psi_{n}} \nn && \times
\int\frac{d\lambda_{\pm}}{2\pi}\int{d\Gamma_{\pm}}
e^{i\lambda_{+}(Y_{+} - \Gamma_{+}) + N_{+}\ln\frac{\Gamma_{+}}{V}
+ ( + \rightarrow - )} . \eqa  Integration over $\lambda_{\pm}$
and $\Gamma_{\pm}$ recovers Eq. (9) in the chiral limit. In the
thermodynamic limit $N_{\pm}, V \rightarrow \infty$ and
${N_{\pm}}/{V}$ fixed, integration over $\Gamma_{\pm}$ and
$\lambda_{\pm}$ can be performed by the saddle point
method\cite{Dmitri}. Integrating over $\Gamma_{\pm}$ first, we
obtain \bqa && Z_{\psi} =
\int\frac{d\lambda_{\pm}}{2\pi}e^{N_{+}\Bigl(\ln\frac{N_{+}}{i\lambda_{+}V}
- 1 \Bigr) + (+\rightarrow -)} \nn && \times
\int{D\psi_{n}}{Da_{\mu}}e^{-\int{d^3x}\bar{\psi}_{n}\gamma_{\mu}(\partial_{\mu}
- ia_{\mu})\psi_{n} + i\lambda_{+}Y_{+} + i\lambda_{-}Y_{-}} .
\eqa An explicit calculation for the instanton average shows that
the vertex $Y^{\pm}$ corresponds to a mass\cite{Dmitri}, $Y^{\pm}
= \int\frac{d^3k}{(2\pi)^{3}}
(2\pi\rho{F}(k))^{2}\bar{\psi}_{n}\frac{1\pm\gamma_{5}}{2}\psi_{n}$
with $\gamma_{5} = \left( \begin{array}{cc} 0 & I \\ -I & 0
\end{array} \right)$\cite{Don_Kim}. Here $F(k)$ is associated with the fermion
zero mode in the effective potential $V^{I}[\psi_{n}]$ in Eq. (8).
In the present paper we do not perform an explicit calculation for
the instanton average in $Y_{\pm}$ and thus we do not know an
exact form of $F(k)$. Our objective is to see how the 't Hooft
interaction appears as an instanton effect. Here $\rho$ is the
size of an instanton. Owing to the neutrality condition of
magnetic charges $N_{+} = N_{-} = N/2$ is obtained in Eq. (11),
where $N$ is the total number of instantons and anti-instantons.
The saddle point solution for $\lambda_{+} = \lambda_{-} \equiv
\lambda$ in Eq. (11) gives rise to the cancellation of the
$\gamma_{5}$ term in the mass, causing the momentum dependent mass
$m(k) = m_{\psi}F^{2}(k)$ with $m_{\psi} =
\lambda(2\pi\rho)^{2}$\cite{Dmitri}. The mass $m_{\psi}$ is
determined by the saddle point equation for $\lambda$ usually
called the self-consistent gap equation\cite{Dmitri} \bqa
&&\frac{8}{N/V}\int\frac{d^3k}{(2\pi)^{3}}\frac{m^{2}(k)}{k^2+m^2(k)}
= 1 . \eqa Ignoring the momentum dependence by setting $F(k) = 1$
for simplicity, we obtain the mass $m_{\psi} =
\frac{\pi}{2\Lambda^{1/2}}\Bigl(\frac{N}{V}\Bigr)^{1/2}$ with the
momentum cut-off $\Lambda$ in small mass limit. Since the mean
density of instantons is proportional to the instanton fugacity,
$N/V \sim y_{m} = e^{- S_{inst}}$ with an instanton action
$S_{inst} \sim 1/e^{2}$\cite{Polyakov,NaLee}, the fermion mass is
roughly given by $m_{\psi} \sim y_{m}^{1/2}$. We obtain the
effective Lagrangian in terms of the Dirac spinon $\psi_{n}$ with
the 't Hooft effective mass $m_{\psi}$ interacting via the
non-compact U(1) gauge field $a_{\mu}$, ${\cal L}_{\psi} =
\sum_{n,m = 1}^{2} \Bigl[
\bar{\psi}_{n}\gamma_{\mu}(\partial_{\mu}\delta_{nm} +
ia_{\mu}\tau^{3}_{nm} )\psi_{m} + m_{\psi}\bar{\psi}_{n}\psi_{n}
\Bigr] $. Despite the mass term we cannot say that the spinon is
really massive. We should show that the instanton fugacity $y_m$
is non-zero. $y_m$ would be determined self-consistently in the
presence of holon contributions.

Including the holon contribution, the non-linear $\sigma$ model
with the easy plane anisotropy and performing a standard duality
transformation\cite{NaLee,Deconfinement,Kim}, we obtain the total
effective Lagrangian in terms of the Dirac spinons and holon
vortices with the electromagnetic field
$A_{\mu}$\cite{Duality_isospin} \bqa && {\cal L} =
\sum_{n=1}^{2}\Bigl[ |(\partial_{\mu} - ic_{n\mu})\Phi_{n}|^{2} +
m^{2}_{\Phi}|\Phi_{n}|^2 + \frac{u_{\Phi}}{2}|\Phi_{n}|^{4} \nn &&
- i(\partial\times{c}_{n})_{\mu}A_{\mu} +
\frac{1}{2\rho}|\partial\times{c}_{n}|^2 \Bigr] \nn && +
\frac{1}{2e^2}|\partial\times{a}|^2 -
i(\partial\times{a})_{\mu}(c_{1\mu} - c_{2\mu}) \nn && + \sum_{n,m
= 1}^{2} \Bigl[
\bar{\psi}_{n}\gamma_{\mu}(\partial_{\mu}\delta_{nm} +
ia_{\mu}\tau^{3}_{nm} )\psi_{m} + m_{\psi}\bar{\psi}_{n}\psi_{n}
\Bigr] . \eqa Here $\Phi_{1(2)}$ is the vortex field with isospin
$\uparrow (\downarrow)$ (isospin $\uparrow (\downarrow)$ meron
field\cite{Deconfinement}) and $c_{1(2)\mu}$, its vortex gauge
field mediating interactions between the vortices. $m_{\Phi}$ and
$u_{\Phi}$ are the mass and self-interaction of the vortices,
respectively. $\rho \sim \delta$ is the coupling strength between
the vortex and vortex gauge field. The presence of the fermion
zero mode in a single instanton potential is the key ingredient
resulting in Eq. (13). In the absence of the fermion zero mode the
term $-y_{m}(\Phi_{1}\Phi_{2}^{\dagger} +
\Phi_{1}^{\dagger}\Phi_{2})$ is usually generated in the dilute
approximation of instantons\cite{NaLee,Deconfinement,Kim}. A
renormalization group ($RG$) study shows instanton condensation
($y_m \rightarrow \infty$)\cite{Kim} inducing vortex pair
condensation $<\Phi_{1}\Phi_{2}^{\dagger}> \not= 0$ in the $SC$
state\cite{NaLee,Kim}. As a result the Higgs-confinement phase
arises\cite{NaLee,Kim}. This state is described by the holon pairs
$z_{1}z_{2}$ [Eq. (2)] in the Higgs field representation. On the
other hand, in the presence of the fermion zero mode this term
makes the partition function zero and thus does not contribute to
our effective Lagrangian Eq. (13). In the dilute approximation of
instantons the fugacity $y_m$ appears only in the fermion mass. At
this level of approximation it is difficult to determine the
instanton fugacity, i.e., the state of instantons. As a matter of
fact it is a long standing unsolved problem to determine the state
of instantons in the presence of matter fields. Generally
speaking, there are two possible instanton states resulting in
deconfined $SC$; one is a dipolar phase ($y_m \rightarrow 0$) and
the other, a "liquid" phase ($0 < y_{m} < \infty$). The latter
does not appear in the Abelian Higgs model\cite{Kim} (without
fermions). But, in the present model we do not have any evidence
to exclude this instanton state. In $(2+1)$ dimensions the basic
trend is confinement, i.e., $y_m \rightarrow
\infty$\cite{Fradkin,NaLee} away from quantum
criticality\cite{U1SL,Deconfinement,Kim}. Owing to the confinement
tendency we should consider dense instantons. But, the presence of
the fermion zero mode does not allow instanton condensation. In
the dense limit a new phase is expected instead of plasma and
dipolar phases. There exist some reports about a new phase in the
two dimensional Coulomb gas when the density of particles is
high\cite{KT}. Furthermore, a new fixed point with non-zero
instanton fugacity was recently reported even in the $QED_3$ with
only massless Dirac fremions\cite{Kleinert}. In this letter we
consider a liquid phase of instantons, i.e., $y_{m} \not= 0$. We
view the emergence of a liquid state as the proximate effect of
the Higgs-confinement phase in the presence of the fermion zero
mode. In the dipolar phase the spinon mass vanishes because of
$m_\psi \sim y_{m}^{1/2} \rightarrow 0$. The resulting effective
field theory is completely the same as Eq. (13) except zero spinon
mass. We do not exclude the possibility of this dipolar phase. We
will discuss this plausible state in a separate publication,
including phase transitions between the three different $SC$
phases.

The mass corresponds to the antiferromagnetic moment of the nodal
fermion\cite{Marston,Don_Kim}. If instantons are in a liquid
state, the $AF$ of the nodal fermions can coexist with the
$d-wave$ $SC$ in underdoped cuprates\cite{Kim_PRL}. The mass can
be considered as an evidence of deconfinement in the underdoped
$SC$ phase. Thus, if the $AF$ is observed in the $SC$ phase,
deconfinement of the spinons and holons is expected to occur. Many
recent experiments have reported the coexistence of the $AF$ and
$SC$\cite{Experiment}. Our new $SC$ may have a chance to be
applicable.

Since the Dirac fermions are massive in the present consideration,
they can be safely integrated out. As a result the Maxwell kinetic
energy ${\cal L}_{a} =
\frac{1}{2\tilde{e}^{2}}|\partial\times{a}|^2$ with $\tilde{e}^{2}
= 12\pi{m_\psi}$\cite{Don_Kim} is generated. Integrating over the
internal gauge field $a_{\mu}$, we obtain a mass term for the
vortex gauge fields, $\frac{e_{eff}^{2}}{2}|c_{1\mu} -
c_{2\mu}|^{2}$ with an effective internal charge $e_{eff}^{2} =
(e^{2}\tilde{e}^2)/(e^{2} + \tilde{e}^2)$. The mass is a relevant
parameter in the $RG$ sense, thus admitting us to set $c_{1\mu} =
c_{2\mu} \equiv c_{\mu}$ in the low energy limit. An effective
Lagrangian is obtained to be in the $SC$ state \bqa && {\cal
L}_{SC} = \sum_{n=1}^{2}\Bigl[ |(\partial_{\mu} -
ic_{\mu})\Phi_{n}|^{2} + m^{2}_{\Phi}|\Phi_{n}|^2 +
\frac{u_{\Phi}}{2}|\Phi_{n}|^{4} \Bigr] \nn && -
i2A_{\mu}(\partial\times{c})_{\mu} +
\frac{1}{\rho}|\partial\times{c}|^2 . \eqa  In the coupling $-
i2A_{\mu}(\partial\times{c})_{\mu}$ an electric charge $2$
originates from both the $z_1$ and $z_2$ bosons. $2e_{el}$
electric charge infers that a vortex quantum is ${hc}/{2e_{el}}$.
Although the underdoped $SC$ state is argued to be the
deconfinement phase, the vortex quantum is not ${hc}/{e_{el}}$ but
${hc}/{2e_{el}}$. This is nothing but the meron-type vortex
discussed in Ref. \cite{SU2_vortex}. The superconductor to
insulator transition induced by the meron vortices is expected to
fall into the XY universality class\cite{Kim_KT,Kim_PRL}. The
above holon vortex Lagrangian is just dual to the non-linear
$\sigma$ model with a non-compact U(1) gauge field in Eq. (1).
This Lagrangian was recently studied by the present author using a
$RG$ analysis\cite{Kim}. In the study the author showed that the
quantum critical point is governed by the XY fixed point. This
result seems to be consistent with experiments for YBCO\cite{XY}.

We would like to comment that the present SU(2) formulation is
valid only in underdoped region\cite{SU2_gauge}. The effect of
SU(2) symmetry breaking may be studied by introducing the Zeeman
terms, $-H_{\psi}\bar{\psi}\tau^{3}\psi$ and
$-H_{z}z^{\dagger}\tau^{3}z$, where $H_{\psi(z)}$ is an effective
"magnetic field" proportional to hole concentration. They would be
important at large doping. $-H_{\psi}\bar{\psi}\tau^{3}\psi$ is
expected to make the fermion zero mode disappear. This can be
checked by investigating the equation of motion [Eq. (3)] in the
presence of the Zeeman term. The role of
$-H_{z}z^{\dagger}\tau^{3}z$ is not clear at present. The
disappearance of the fermion zero mode will cause the
Higgs-confinement phase described by Eq. (2). We anticipate a
quantum phase transition between the deconfined $SC$ [Eq. (13)]
and the $BCS$ one [Eq. (2)] at some critical doping inside the
$SC$ dome. This interesting possibility will be studied near
future.

K.-S. Kim especially thanks Dr. Yee, Ho-Ung for helpful
discussions of the fermion zero mode and 't Hooft interaction.


\begin{thebibliography}{9}
\bibitem{Senthil_Lee}  T. Senthil and P. A. Lee,
cond-mat/0406066.
\bibitem{U1SL} M. Hermele et al., Phys. Rev. B {\bf 70}, 214437 (2004).
\bibitem{SU2_gauge} P. A. Lee et al., Phys. Rev. B {\bf 57}, 6003 (1998); P. A. Lee and N. Nagaosa,
Phys. Rev. B {\bf 68}, 024516 (2003).
\bibitem{Jackiw} R. Jackiw and C. Rebbi, Phys. Rev. D {\bf 13},
3398 (1976).
\bibitem{tHooft} G. 't Hooft, Phys. Rev. D {\bf 14}, 3432 (1976).
\bibitem{Dmitri} D. Diakonov, Lectures at the Enrico Fermi
School in Physics, Varenna, (1995); hep-ph/9602375.
\bibitem{Don_Kim} D. H. Kim and P. A. Lee, Annals Phys. {\bf{ 272}}, 130
(1999).
\bibitem{Marston} J. B. Marston, Phys. Rev. Lett. {\bf 64}, 1166 (1990).
\bibitem{Compactness} The kinetic energy is given by
$-\frac{1}{e^2}\sum_{n}\cos(\partial\times{a})_{n}$ in lattice,
where $n$ is its dual lattice. We approximate this to the Maxwell
form in the continuum limit, keeping the compactness of the gauge
field.
\bibitem{Fradkin} E. Fradkin and S. H. Shenker, Phys. Rev. D {\bf 19}, 3682 (1979).
\bibitem{DHLee} D. H. Lee, Phys. Rev. Lett {\bf 84}, 2694 (2000).
\bibitem{Kim_KT} K.-S. Kim et al., Phys. Rev. B {\bf 69}, 014504 (2004).
\bibitem{Instanton_monopole} An instanton solution representing a
tunnelling event between energetically degenerate but
topologically inequivalent gauge vacua is nothing but a magnetic
monopole in two space dimensions and one imaginary time
dimension\cite{Polyakov}.
\bibitem{Potential} L. H. Ryder, Quantum Field Theory (2nd., Cambridge University Press,
1996).
\bibitem{Polyakov} A. M. Polyakov, Gauge Fields and Strings (ch.4), harwood academic publishers. (1987).
\bibitem{Log_confinement} Even in the deconfinement phase the Coulomb
interaction is logarithmically confining in two space dimensions.
\bibitem{NaLee} N. Nagaosa and P. A. Lee, Phys. Rev. B {\bf 61}, 9166
(2000).
\bibitem{Deconfinement} T. Senthil et al., Science {\bf 303}, 1490 (2004);
T. Senthil et al., Phys. Rev. B {\bf 70}, 144407 (2004).
\bibitem{Kim} K.-S. Kim, cond-mat/0406511.
\bibitem{Duality_isospin} The isospin interaction is expected to
be marginally irrelevant in the $RG$ sense. Thus it does not
affect the $SC$ transition. It can be ignored in the absence of
the single instanton excitations.
\bibitem{KT} J.-R. Lee and S. Teitel, Phys. Rev. B {\bf 46}, 3247
(1992); P. Gupta and S. Teitel, Phys. Rev. B {\bf 55}, 2756
(1997); The new phase is the ionic lattice state. Although it is
not a liquid phase owing to different Coulomb potentials, we have
a clue to expect another new phase.
\bibitem{Kleinert} F. S. Nogueira and H. Kleinert,
cond-mat/0501022.
\bibitem{Kim_PRL}  K.-S. Kim et al., cond-mat/0404527.
\bibitem{Experiment} J. E. Sonier et al., Science {\bf 292}, 1692 (2001);
Y. Sidis et al., Phys. Rev. Lett. {\bf 86}, 4100 (2001); H. A.
Mook et al., Phys. Rev. B {\bf 64}, 012502 (2001); M. -H. Julien
et al., Phys. Rev. B {\bf 63}, 144508 (2001); S. Ono et al., Phys.
Rev. Lett. {\bf 85}, 638 (2000); Ch. Niedermayer et al., Phys.
Rev. Lett. {\bf 80}, 3843 (1998).
\bibitem{SU2_vortex} P. A. Lee and X.-G. Wen, Phys. Rev. B {\bf 63}, 224517
(2001).
\bibitem{XY} F. S. Nogueira, Phys. Rev. B {\bf 62}, 14559 (2000); references therein; D. J.
Lee and I. D. Lawrie, Phys. Rev. B {\bf 64}, 184506 (2001);
references therein.
\end{thebibliography}
\end{document}